\documentclass[a4paper]{jpconf}
\usepackage{graphicx}
\usepackage{dcolumn}
\usepackage{bm}
\usepackage{stackrel}
\usepackage{amsmath}
\usepackage{amsfonts}
\usepackage{amssymb}
\usepackage{amsthm}
\usepackage{epstopdf}
\usepackage{longtable}
\usepackage{array}
\usepackage{pifont}
\usepackage{latexsym}
\usepackage[latin1]{inputenc}
\usepackage{bm}
\usepackage{textcomp}
\usepackage{subfigure}
\usepackage{braket}
\usepackage{wasysym}
\usepackage{ulem}
\usepackage{color, soul}
\usepackage{graphicx}
\begin{document}
\title{Composite pulses in Hyper-Ramsey spectroscopy for the next generation of atomic clocks}

\author{T. Zanon-Willette$^{1}$, M. Minissale$^{2,3}$, V.I. Yudin$^{4}$ and A.V. Taichenachev$^{4}$}
\address{$^{1}$ LERMA, Observatoire de Paris, PSL Research University, CNRS, Sorbonne Universités, UPMC Univ. Paris 06, F-75005, Paris, France}
\address{$^{2}$ Aix Marseille Université, CNRS, PIIM UMR 7345, 13397 Marseille, France}
\address{$^{3}$ Aix-Marseille Université, CNRS, Centrale Marseille, Institut Fresnel UMR 7249, 13013 Marseille, France}
\address{$^{4}$ Institute of Laser Physics, Siberian Branch of Russian Academy of Sciences, Novosibirsk 630090, Russia,
Novosibirsk State University, Novosibirsk 630090, Russia,
and Novosibirsk State Technical University, Novosibirsk 630092, Russia}

\ead{thomas.zanon@upmc.fr}

\begin{abstract}
The next generation of atomic frequency standards based on an ensemble of neutral atoms or a single-ion will provide very stringent tests in metrology, applied and fundamental physics requiring a new step in very precise control of external systematic corrections.
In the proceedings of the 8th Symposium on Frequency Standards and Metrology, we present a generalization of the recent Hyper-Ramsey spectroscopy with separated oscillating fields using composites pulses in order to suppress field frequency shifts induced by the interrogation laser itself. Sequences of laser pulses including specific selection of phases, frequency detunings and durations are elaborated to generate spectroscopic signals with a strong reduction of the light-shift perturbation by off resonant states. New optical clocks based on weakly allowed or completely forbidden transitions in atoms, ions, molecules and nuclei will benefit from these generalized Ramsey schemes to reach relative accuracies well below the 10$^{-18}$ level.
\end{abstract}

\section{Introduction}

\begin{figure}[t!!]
\centering%
\resizebox{12.7cm}{!}{\includegraphics[angle=-90]{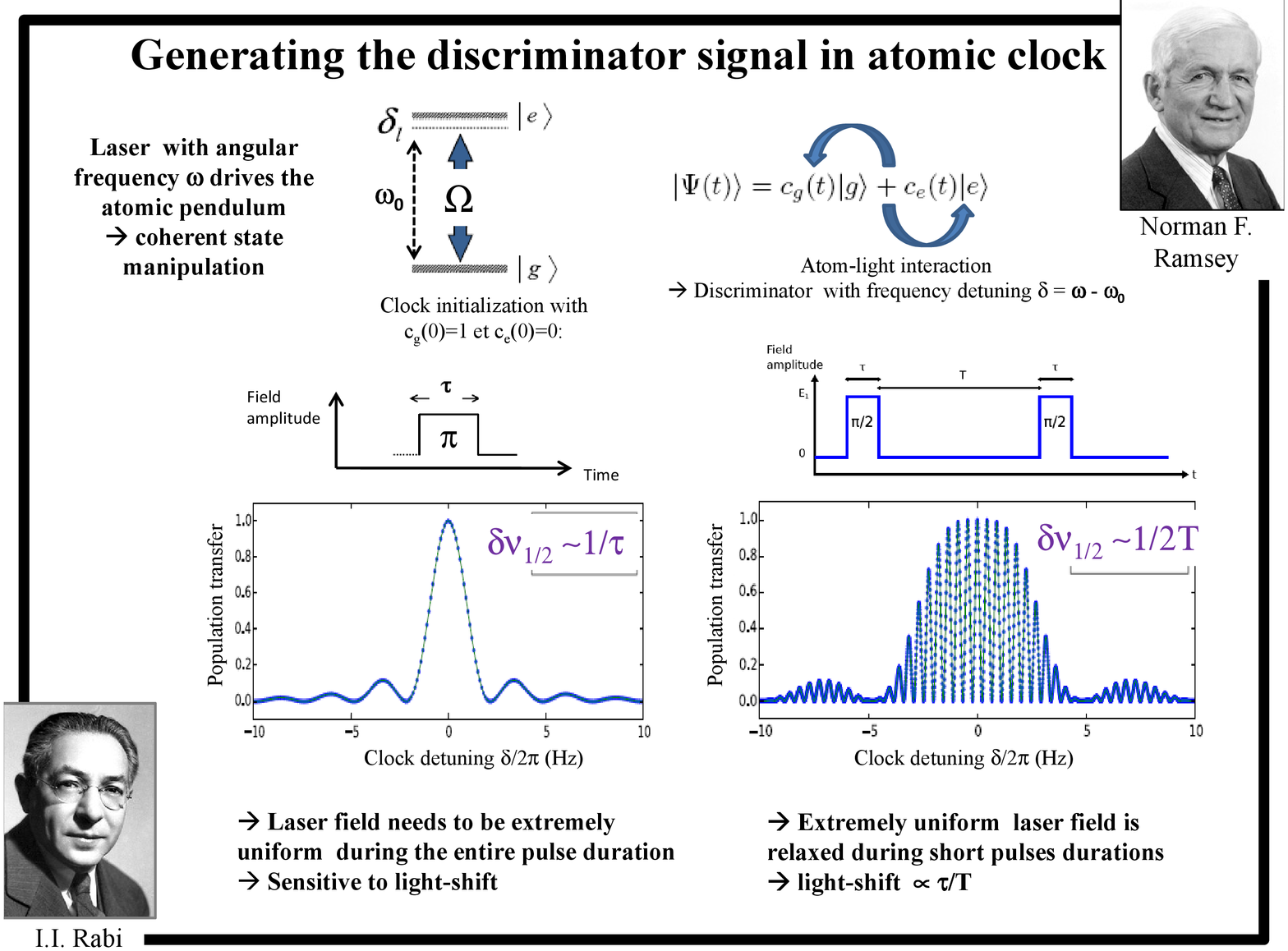}}
\caption{(color online). Quantum system with a narrow $|g\rangle\leftrightarrow|e\rangle$ clock transition probed by a single laser pulse excitation. The unperturbed clock detuning is introduced by $\delta$ and the Rabi frequency is $\Omega$. The usual pulse area is defined by the product $\Omega\tau$. Rabi and Ramsey spectroscopy are two laser probing techniques generating a frequency discriminator for an atomic clock. The full width at half maximum is defined by $\delta\nu_{1/2}$.}
\label{Figure-1}
\end{figure}

\indent The story of very high precision laser pulsed spectroscopy has started in 1930 with the atomic and molecular beam resonance method invented by I.I. Rabi to improve the resolution of frequency measurements \cite{Rabi:1938}. Under a quasi-resonant irradiation, a quantum two-level system can undergo coherent Rabi oscillations. In this case, the initial two-level occupation probabilities can be inverted by a so-called $\pi$ pulse with an adequate selection of laser field frequency and pulse duration. By scanning the frequency of the irradiation around the exact resonance, a narrow spectroscopic resonance is observed that can be used to realize a very precise discriminator stabilizing a local frequency oscillator to the atomic or molecular transition.
To improve even further the frequency resolution, N.F. Ramsey has proposed to replace the single oscillatory field by a double microwave excitation pulse with $\pi/2$ pulses separated by a free evolution time \cite{Ramsey:1950} as schematized in Fig.~\ref{Figure-1}. The main advantage of the two-pulse Ramsey spectroscopy is to reduce the probing electromagnetic field perturbation on the atomic transition itself which has drastically impacted the time and frequency metrology with microwave atomic clocks since the 1950s \cite{Vanier:1989}.

Optical frequency-standards are today actively developed to replace microwave transitions by very narrow optical resonances in single ion, fermionic and bosonic alkaline-earth atoms \cite{Ludlow:2015}. Very long storage time of Doppler and recoil free quantum particles have been achieved using laser cooling techniques and the level of 10$^{-18}$ relative accuracy will be probably almost achieved in the near future \cite{Derevianko:2011}. As an example, the lattice fermionic clock requires now a very precise control of atomic interactions to cancel some systematic frequency shifts below the relative $10^{-17}$ level of accuracy \cite{Hinkley:2013,Nicholson:2015}.
Light-shifts represent a non negligeable issue for clocks based on bosonic neutral atoms with forbidden dipole transitions activated by mixing static magnetic field with a single laser \cite{Taichenachev:2006,Barber:2006}, for clocks with magic-wave induced transition in even isotopes \cite{Ovsiannikov:2007} or with a E1-M1 two-photon laser excitation \cite{Santra:2005,Zanon:2006,Zanon:2014}. Because these important issues have to be addressed for bosons, we present in these proceedings a generalization of the Ramsey spectroscopy to reduce at a very low level of correction the light-shift induced by the probing laser itself on the atomic resonance.

\section{Clock frequency shift in a generalized Ramsey two-pulse spectroscopy}

\begin{figure}[b!!]
\centering%
\resizebox{12.7cm}{!}{\includegraphics[angle=-90]{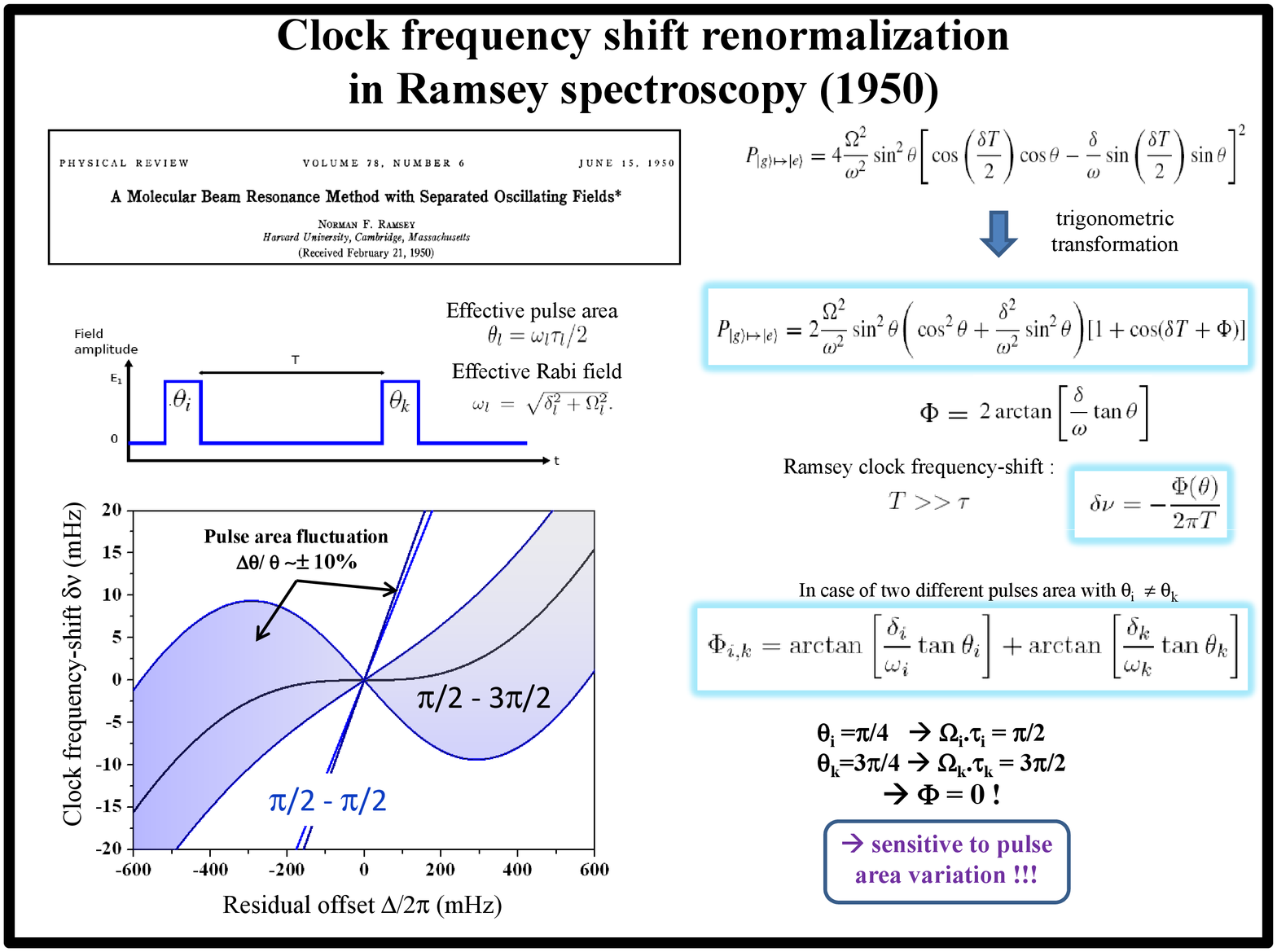}}
\caption{(color online). Clock frequency shift $\delta\nu$ affecting the central fringe in a Ramsey two-pulse spectroscopy. The central fringe Ramsey frequency shift $\delta\nu=-\Phi_{R}/2\pi T$ with $\pi/2-\textup{T}-\pi/2$ pulses is compared to the Hyper-Ramsey frequency shift $\delta\nu=-\Phi_{HR}/2\pi T$ applying a $\pi/2-\textup{T}-3\pi/2$ configuration. The shadow area between lines corresponds to an uncompensated $\pm10\%$ pulse area variation. The Rabi frequency is defined as $\Omega=\pi/2\tau$ where $\tau=0.1875$~s and T$=2$~s.}
\label{Figure-2}
\end{figure}

\indent In a Ramsey laser spectroscopy defined by two pulses with effective area $\theta_{l}$ ($l=\textup{i,k}$),
the single Rabi's $\pi$ pulse interaction zone is splitting into two very short $\pi/2$ pulses with duration $\tau$ separated by a free evolution time T. The atoms spend a much shorter time in the presence of the external electromagnetic fields minimizing inhomogeneities which may reduce precision in determining the transition frequency. When we consider a light-shift effect during the laser pulses, Ramsey spectroscopy renormalizes the light perturbation by the factor $\tau/T$. If Ramsey fringes are largely shifted by off-resonant states, an additional laser frequency step may be applied to rectify the anticipated shift, thus requiring a redefinition of frequency detunings as $\delta_{l}=\delta+\Delta_{l}$.
The uncompensated residual contribution $\Delta_{l}$ present during pulses leads to a net clock frequency-shift of the central fringe expressed as:
\begin{equation}
\begin{split}
\delta\nu=-\frac{\Phi(\delta\rightarrow0,\Delta_{l})}{2\pi\textup{T}}
\end{split}
\label{clock-frequency-shift}
\end{equation}
where $\Phi(\delta\rightarrow0,\Delta_{l})$ is the zero order phase-shift affecting the central fringe when the unperturbed clock detuning $\delta$ between the laser frequency and the atomic transition is closed to zero. If we simply consider the extension of the two-pulse configuration, as in the initial Hyper-Ramsey scheme with different pulse areas \cite{Yudin:2010}, we found the global phase-shift expression as \cite{Zanon:2014}:
\begin{equation}
\begin{split}
\Phi_{HR}=\arctan\left[\frac{\frac{\delta_{\textup{i}}}{\omega_{\textup{i}}}\tan\theta_{\textup{i}}+\frac{\delta_{\textup{k}}}{\omega_{\textup{k}}}\tan\theta_{\textup{k}}}{1-\frac{\delta_{\textup{i}}\delta_{\textup{k}}}{\omega_{\textup{i}}
\omega_{\textup{k}}}\tan\theta_{\textup{i}}\tan\theta_{\textup{k}}}\right]
\end{split}
\label{Hyper-Ramsey-phase}
\end{equation}
where the effective area is $\small{\theta_{l}=\omega_{l}\tau_{l}/2}$ and $\small{\omega_{l}=\sqrt{\delta_{l}^{2}+\Omega_{l}^{2}}}$. The uncompensated part of the residual light-shift correction becomes proportional to the laser pulse area and is reduced to usual Ramsey phase-shift $\small{\Phi_{R}=2\arctan\left[\frac{\Delta}{\omega}\tan\theta\right]}$ where $\small{\omega=\sqrt{\Delta^{2}+\Omega^{2}}}$ in Eq.~\ref{Hyper-Ramsey-phase} when $\theta_{\textup{i}}=\theta_{\textup{k}}=\theta$ and $\delta_{\textup{i}}=\delta_{\textup{k}}\equiv\Delta$. Such a linear dependence with $\pi/2$ pulses versus the uncompensated part of the light perturbation is shown by the solid blue line in Fig.~\ref{Figure-2}. The frequency shift response is usually observed in microwave atomic clocks using Ramsey spectroscopy \cite{Vanier:1989}.
\begin{figure}[t!!]
\centering%
\resizebox{9.0cm}{!}{\includegraphics[angle=-90]{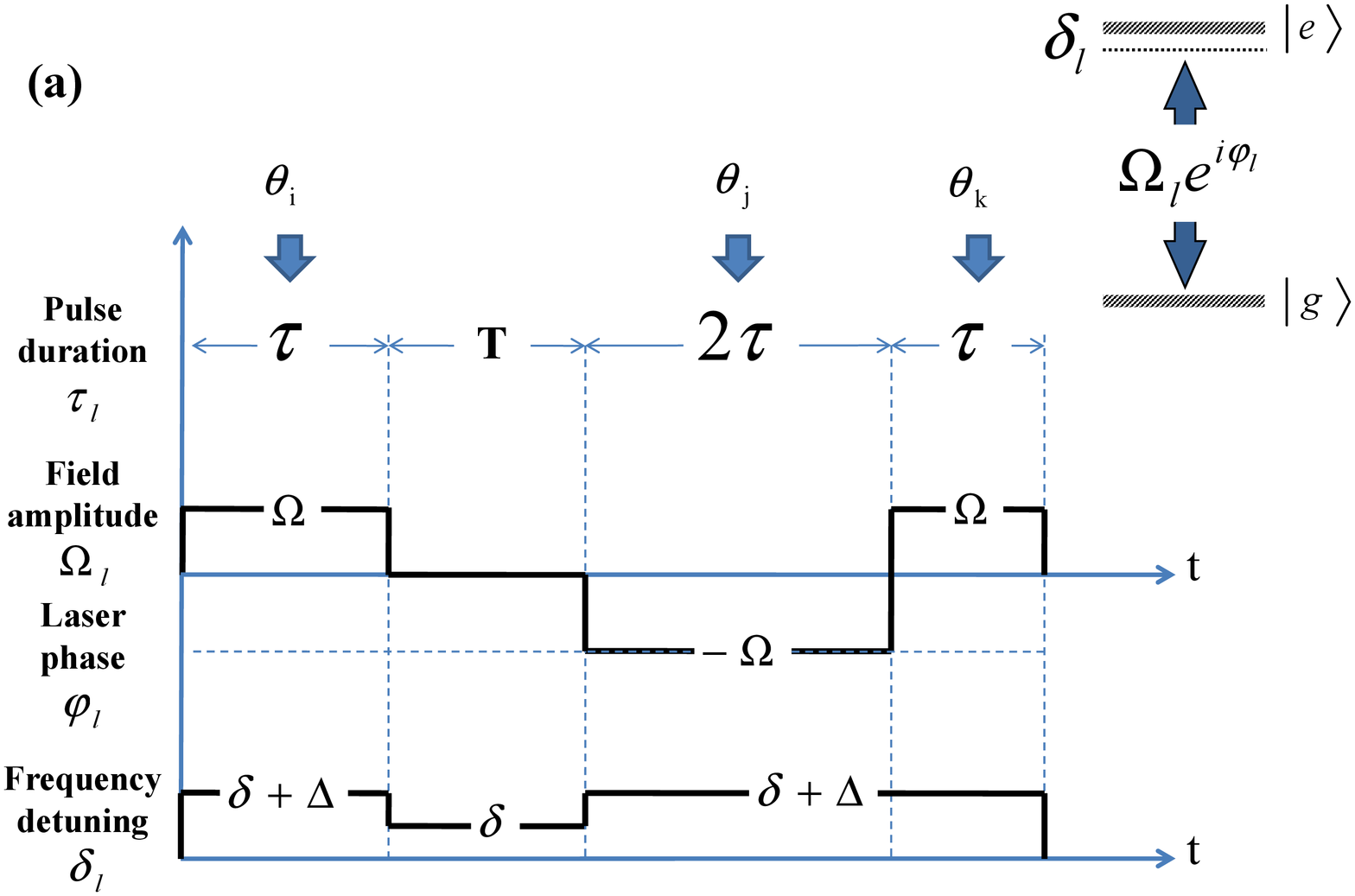}}\resizebox{8.0cm}{!}{\includegraphics[angle=-90]{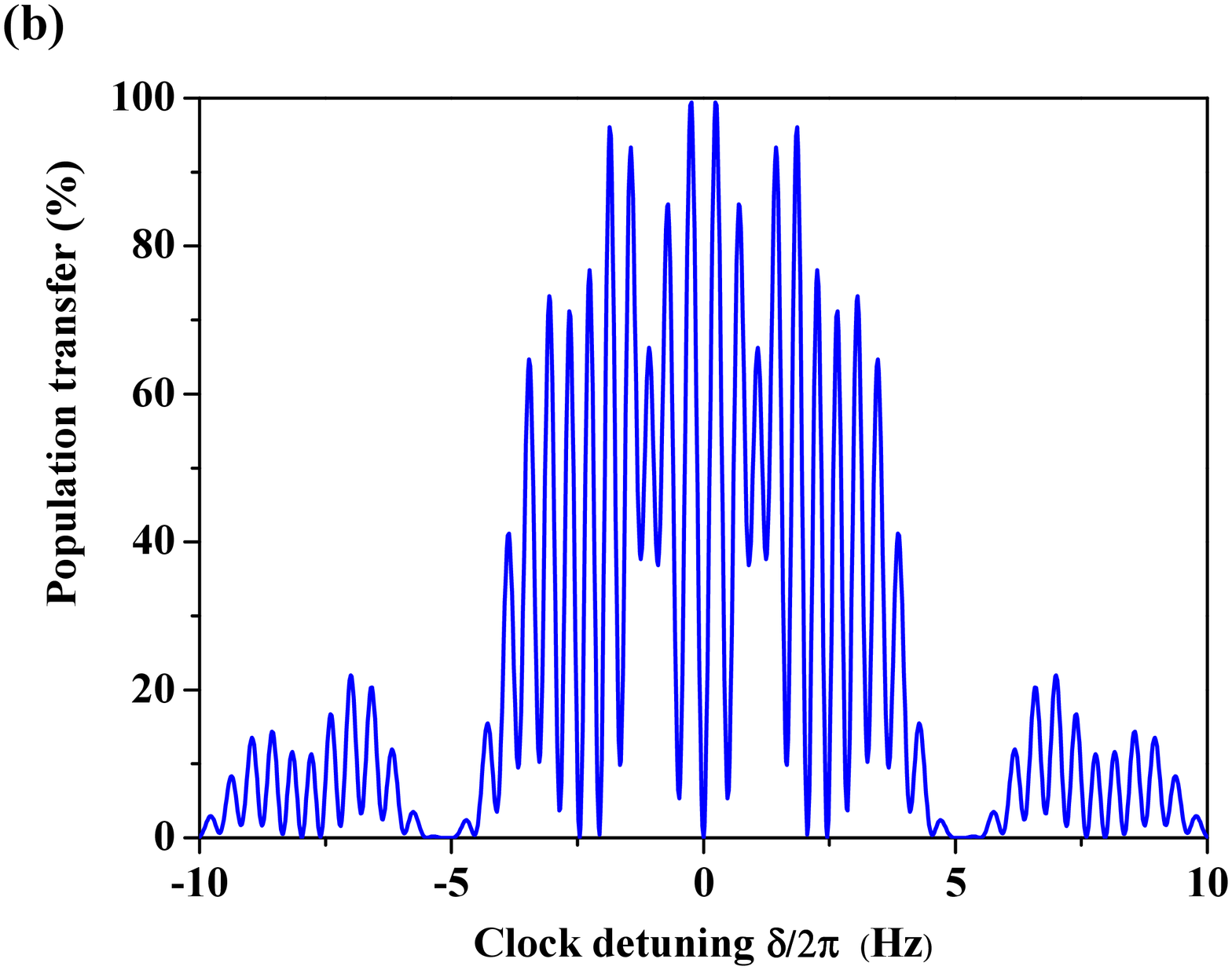}}
\caption{(color online). (a) Sequence of composite pulses $\theta_{\textup{i}}-\textup{T}-\theta_{\textup{j}},\theta_{\textup{k}}$ with specified laser parameters including detuning $\delta_{l}$, complex field amplitude $\Omega_{l}e^{i\varphi_{l}}$, pulse duration $\tau_{l}$  where $l=\textup{i,j,k}$ and a free evolution time T between pulses. (b) Generalized Hyper-Ramsey resonance generated by composite pulses.}
\label{Figure-3}
\end{figure}
More importantly, there is another pulse configuration which was never studied by Ramsey to strongly reduce the frequency-shift induced by the laser probe itself.
By selecting in Eq.~\ref{Hyper-Ramsey-phase} effective pulse areas such that $\theta_{\textup{i}}=\theta$ and $\theta_{\textup{k}}=3\theta$ taking for example a pulse duration $\tau$ three times longer during the final pulse, we observe a nonlinear clock frequency-shift of the central Ramsey fringe versus the residual correction $\Delta$ shown by the solid black line in Fig.~\ref{Figure-2}.
The non standard reduction of the residual light-shift perturbation is produced by a combination of pulse areas $\theta\sim\pi/4~(\Omega\tau=\pi/2)$ and $\theta\sim3\pi/4~(\Omega\tau=3\pi/2)$ in Eq.~\ref{Hyper-Ramsey-phase}. The linear dependence of the clock frequency shift on small variations of the light-shifts is totally removed except for a residual cubic correction originated by the phase of Eq.~\ref{Hyper-Ramsey-phase}.
However, a relative pulse area modification by $\pm10\%$ is still affecting the result if it is not well compensated.

\begin{figure}[b!!]
\centering%
\resizebox{12.7cm}{!}{\includegraphics[angle=-90]{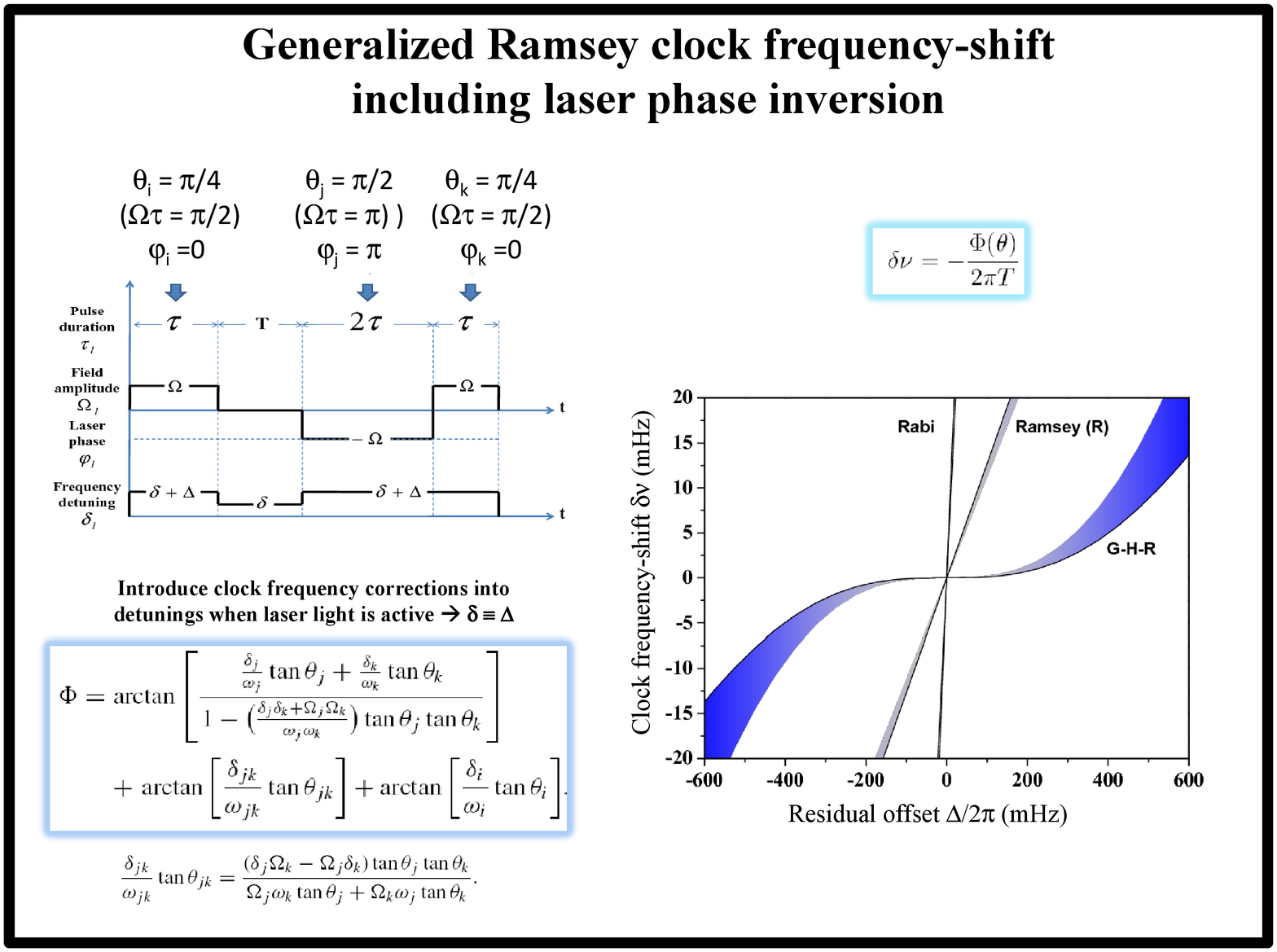}}
\caption{(color online). Generalized Hyper-Ramsey frequency shift $-\Phi_{GHR}/2\pi T$ versus a residual uncompensated light-shift correction. The sequence of composite pulses with $\pi/2-\textup{T}-\pi,\pi/2$ pulses (including $\varphi_{\textup{j}}=\pi$ during the intermediate $\pi$-pulse) is capable of minimizing the pulse area variation to realize a robust frequency discriminator against a residual light-shift perturbation from the probing laser itself.}
\label{Figure-4}
\end{figure}

\section{Generalized Hyper-Ramsey clock frequency shift with composite laser pulses}

\indent Generalization of the Ramsey scheme from the original two-pulse configuration \cite{Ramsey:1950} to more complex ones including additional intermediate pulses permits to reduce shifts and broadening due to inhomogeneous excitation conditions or shifts that are the result of the excitation itself \cite{Hahn:1950,Levitt:1986}. Sequence of three excitation pulses $\theta_{l}$ ($l=\textup{i,j,k}$) shown in Fig.~\ref{Figure-3}(a) with suitably selected frequency $\delta_{l}$ and phase steps $\varphi_{l}$ have been introduced in order to reduce the light-shift and to efficiently suppress the sensitivity of the spectroscopic signal to variations of the probe light intensity \cite{Yudin:2010,Zanon:2015}.
These new schemes, based on the Generalized Hyper-Ramsey (GHR) resonance reported in Fig.~\ref{Figure-3}(b), have been tested experimentally for few clock transitions as the ultra-narrow electric octupole transition in the $^{171}$Yb$^{+}$ clock \cite{Huntemann:2012}.
Composite pulses inspired by NMR techniques are used to compensate simultaneously for noise decoherence, pulse area fluctuation and residual frequency offset due to the applied laser field itself. The first or the second pulse of the usual Ramsey sequence can be separated in two or more contiguous sections which yield to new efficient probing protocols with more degrees of freedom to minimize resonance shifts.
The generalized Hyper-Ramsey phase-shift accumulated after a three-pulse interrogation scheme becomes \cite{Zanon:2015}:
\begin{equation}
\begin{split}
\Phi_{GHR}=&\arctan\left[\frac{\frac{\frac{\delta_{\textup{j}}}{\omega_{\textup{j}}}\tan\theta_{\textup{j}}+\frac{\delta_{\textup{k}}}{\omega_{\textup{k}}}\tan\theta_{\textup{k}}}{1-\frac{\delta_{\textup{j}}\delta_{\textup{k}}+\Omega_{\textup{j}}\Omega_{\textup{k}}}{\omega_{\textup{j}}\omega _{\textup{k}}}\tan\theta_{\textup{j}}\tan\theta_{\textup{k}}}+\frac{\frac{\delta_{\textup{i}}}{\omega_{\textup{i}}}\tan\theta_{\textup{i}}+\frac{\delta_{\textup{jk}}}{\omega_{\textup{jk}}}\tan\theta_{\textup{jk}}}{1-\frac{\delta_{\textup{i}}\delta_{\textup{jk}}}{\omega_{\textup{i}}\omega_{\textup{jk}}}\tan\theta_{\textup{i}}\tan\theta_{\textup{jk}}}}{1-\frac{\frac{\delta_{\textup{j}}}{\omega_{\textup{j}}}\tan\theta_{\textup{j}}+\frac{\delta_{\textup{k}}}{\omega_{\textup{k}}}\tan\theta_{\textup{k}}}{1-\frac{\delta_{\textup{j}}\delta_{\textup{k}}+\Omega_{\textup{j}}\Omega_{\textup{k}}}{\omega_{\textup{j}}\omega _{\textup{k}}}\tan\theta_{\textup{j}}\tan\theta_{\textup{k}}}\frac{\frac{\delta_{\textup{i}}}{\omega_{\textup{i}}}\tan\theta_{\textup{i}}+\frac{\delta_{\textup{jk}}}{\omega_{\textup{jk}}}\tan\theta_{\textup{jk}}}{1-\frac{\delta_{\textup{i}}\delta_{\textup{jk}}}{\omega_{\textup{i}}\omega_{\textup{jk}}}\tan\theta_{\textup{i}}\tan\theta_{\textup{jk}}}}\right]
\end{split}
\label{Generalized-Hyper-Ramsey-phase}
\end{equation}
with a reduced variable
\begin{equation}
\small{
\begin{split}
\frac{\delta_{\textup{jk}}}{\omega_{\textup{jk}}}\tan\theta_{\textup{jk}}=\frac{\left(\delta_{\textup{j}}\Omega_{\textup{k}}-\Omega_{\textup{j}}\delta_{\textup{k}}\right)\tan\theta_{\textup{j}}\tan\theta_{\textup{k}}}{\Omega_{\textup{j}}\omega_{\textup{k}}\tan\theta_{\textup{j}}+\Omega_{\textup{k}}\omega_{\textup{j}}\tan\theta_{\textup{k}}}
\end{split}}
\label{reduced-variable}
\end{equation}
The key point is to establish some efficient quantum control protocols which compensate for frequency-shift. Moreover the quantum control protocols have to be robust to small change in pulse area while achieving a highly contrasted population transfer between the targeted states \cite{Yudin:2010,Zanon:2015}. By imposing specific pulse areas in Eq.~\ref{Generalized-Hyper-Ramsey-phase} as $\theta_{\textup{i}}=\theta_{\textup{k}}=\theta$ ($\Omega\tau=\pi/2$) and $\theta_{\textup{j}}=2\theta$ ($\Omega\tau=\pi$) while fixing laser field phases at $\varphi_{\textup{i}}=\varphi_{\textup{k}}=0$ and $\varphi_{\textup{j}}=\pi$ ($\Omega\rightarrow-\Omega$) during the intermediate pulse, a very good compensation of the $\pm10\%$ relative pulse area variation is demonstrated as reported in Fig.~\ref{Figure-4}.

It seems that combinations of generalized Hyper-Ramsey spectral resonances including some phase-step controls and different pulse durations may demonstrate in the future their capacity to eliminate systematic corrections including potential biases or higher-order light-shift contributions from the probe laser.
The next generation of optical clocks based on weakly allowed or completely forbidden transitions in atoms, ions, molecules and nuclei will benefit from these generalized Hyper-Ramsey schemes to reach relative accuracies lower than the 10$^{-18}$ level.
The Generalized Hyper-Ramsey resonance with separated oscillating fields has also been extended to bosonic three-levels systems using optical stimulated Raman transitions \cite{Zanon:2014} and some recent modified Hyper-Ramsey spectroscopy schemes based on various phase-step modulations of laser pulses have been proposed to completely eliminate the light-shift in optical clocks \cite{Hobson:2016,Zanon-2:2015,Yudin:2016}.

\subsection*{\textup{\textbf{Acknowledgments}}}
\indent T. Zanon-Willette deeply acknowledges the organizers of the 8th symposium on frequency standards and metrology. The related proceedings are based on the material presented during the talk delivered on the 15th of october 2015.

V.I.Yu. and A.V.T. were supported by the RFBR (grants 14-02-00712, 14-02-00939, 14-02-00806), by the Russian Academy of Sciences, by Presidium of the Siberian Branch of the Russian Academy of Sciences, by the RF Ministry of Education and Science (state assignment No. 2014/139 project No. 825).

\end{document}